\documentclass[11pt]{article} 

\usepackage{jheppub}

\usepackage{amssymb,amsbsy,amsfonts,amsthm}
\usepackage{bbm}
\usepackage{cancel}
\usepackage[normalem]{ulem}
\usepackage[capitalise]{cleveref}

\usepackage{makecell}

\setcellgapes{2pt}

\newcommand{\bfr}{{\mathbf{r}}}

\usepackage{youngtab}
\newcommand\tyng[1]{{\tiny\yng(#1)}}
\usepackage{tikz}
\usetikzlibrary{calc,shapes,fit,positioning,decorations.pathreplacing,calligraphy,arrows.meta}
\tikzset{every picture/.style={remember picture}}

\newcommand{\U}{\operatorname{U}}
\newcommand{\SU}{\operatorname{SU}}
\renewcommand{\O}{\operatorname{O}}
\newcommand{\SO}{\operatorname{SO}}
\newcommand{\Sp}{\operatorname{Sp}}
\newcommand{\so}{\mathfrak{so}}
\renewcommand{\sp}{\mathfrak{sp}}
\newcommand{\e}{\operatorname{e}}
\renewcommand{\d}{\mathrm{d}}
\renewcommand{\i}{\mathrm{i}}
\newcommand{\tr}{\operatorname{tr}}

\newcommand{\sgn}{\operatorname{sgn}}
\newcommand{\adj}{\mathrm{adj}}
\renewcommand\>{\rangle}
\newcommand\<{\langle}

\theoremstyle{definition}
\theoremstyle{plain}
\newtheorem{Theorem}{Theorem}

\renewcommand\sc[1]{\textnormal{\color{red!50!black}{[sc: #1]}}}

\title{Qubit Regularization and Qubit Embedding Algebras}

\author[1]{Hanqing Liu}
\author[1]{and Shailesh Chandrasekharan}

\affiliation[1]{Department of Physics, Duke University, Durham NC 27708-0305} 
\emailAdd{hanqing.liu@duke.edu} \emailAdd{sch27@duke.edu}

\abstract{Qubit regularization is a procedure to regularize the infinite dimensional local Hilbert space of bosonic fields to a finite dimensional one, which is a crucial step when trying to simulate lattice quantum field theories on a quantum computer. When the qubit-regularized lattice quantum fields preserve important symmetries of the original theory, qubit regularization naturally enforces certain algebraic structures on these quantum fields. We introduce the concept of qubit embedding algebras (QEAs) to characterize this algebraic structure associated with a qubit regularization scheme. We show a systematic procedure to derive QEAs for the $\O(N)$ lattice spin models and the $\SU(N)$ lattice gauge theories. While some of the QEAs we find were discovered earlier in the context of the D-theory approach, our method shows that QEAs are far more richer. A more complete understanding of the QEAs could be helpful in recovering the fixed points of the desired quantum field theories.}


\begin{document}

\maketitle
\flushbottom
\section{Introduction}

Recent developments in quantum technologies is making the dream of quantum computing into a realistic and exciting possibility \cite{Ladd_2010}. This has triggered the hope that we can overcome the sign problem that plague Monte Carlo calculations of quantum systems \cite{Troyer:2004ge}. The sign problem arises essentially due to the highly entangled nature of quantum states \cite{Preskill:2018fag} and currently limits our ability to understand real time evolution of quantum systems and to compute the ground state properties of fermionic matters. Overcoming these challenges using a quantum computer could help us achieve a much deeper understanding of many fundamental laws of the nature. Quantum simulations of physical systems using simple models \cite{Jordan:2012xnu,Jordan:2011ci,Jordan:2014tma,Jordan:2017lea,Dumitrescu:2018njn,Lu:2018pjk,Kokail:2018eiw,Klco:2018kyo,Klco:2018zqz,Lamm:2018siq} and both abelian \cite{Banerjee:2012pg,Bazavov:2015kka,Unmuth-Yockey:2018xak,Meurice:2019ddf} and non-abelian \cite{Mathur:2004kr,Anishetty:2009nh,Raychowdhury:2018osk,Raychowdhury:2019iki,Davoudi:2020yln,Raychowdhury:2021jbo} lattice gauge theories \cite{Banuls:2019bmf} have already been proposed using both analog and digital quantum simulators, such as ultracold atoms in optical lattices \cite{Jaksch:1998zz,Tagliacozzo:2012vg,Tagliacozzo:2012df,Zohar:2012xf,Zohar:2013zla,Zohar:2015hwa,Dasgupta:2020itb}, trapped ions \cite{Cirac:1995zz,Schindler:2013,Martinez:2016yna} and superconducting circuits \cite{You:2005,Marcos:2014lda,Mezzacapo:2015bra}. 

Simulating quantum field theories using a quantum computer creates its own challenges. Calculations of quantities in quantum field theories are often plagued with infinities and a regularization procedure is necessary to even define the calculation. The word regularization usually refers to a method that removes the ultraviolet infinities that arise due to the presence of an infinite number of quantum degrees of freedom in any ``small'' spatial region. In the Hamiltonian formulation, a spatial lattice often provides a non-perturbative regularization of these ultraviolet infinities. A quantum critical point in the lattice theory provides a way to remove the regularization and define the continuum quantum field theory. This is due to the fact that the long distance physics of the model in the vicinity of the correct quantum critical point, is described by the RG fixed point that describes the quantum field theory.

In bosonic quantum field theories there is yet another type of infinity, i.e., the infinite dimensional Hilbert space at every local lattice site. Local lattice bosonic field $\phi_\bfr$ and its conjugate field $\pi_\bfr$ on spatial sites $\bfr$ satisfy the canonical commutation relation
\begin{align}
  [\phi_\bfr,\pi_{\bfr'}] \ =\ \i \delta_{\bfr,\bfr'},
  \label{eq:ccrel}
\end{align}
which can only be realized with an infinite dimensional local Hilbert space. Although this infinity does not make calculations singular, it does create challenges for simulating quantum field theories on a quantum computer, whose local Hilbert space is finite dimensional. The procedure of regularizing this infinite dimensional local Hilbert space to a finite dimensional one is what we call ``qubit regularization'', which has been proposed as a first step in simulating quantum field theories using a quantum computer \cite{PhysRevLett.123.090501,Singh:2019uwd,Singh:2019jog,Bhattacharya:2020gpm,Zhou:2021qpm}.






In analogy with ultraviolet regularization, qubit regularization can also be accomplished in many ways. In fact all lattice quantum spin models, that have been studied in condensed matter physics over the years, can be considered as examples of qubit regularized models due to their finite dimensional local Hilbert space. 
Lattice gauge theories were also brought under this framework through the quantum link formulations many years ago \cite{Chandrasekharan:1996ih,Brower:1997ha,Brower:2003vy}, and quantum simulation based on the quantum link models is also studied \cite{Banerjee:2012xg,Wiese:2013uua,Wiese:2014rla}. Recently, discretizing the continuous symmetries has also been proposed \cite{Bender:2018rdp,Hackett:2018cel,Lamm:2019bik,Alexandru:2019nsa}. Given the limited computational resources in the NISQ era \cite{Preskill:2018}, it is crucial to explore the best qubit regularization schemes that can use the limited resources in a clever way. Here we believe that preserving the symmetries of the original theory should be an important criterion during the regularization procedure, since it may help in constructing qubit Hamiltonians within the basin of attraction of the correct RG fixed point with the same symmetry. 
Besides, it is also desirable to understand the properties of the regularized Hilbert space and learn if there are interesting hidden mathematical structures that characterize the regularization scheme. 

In the qubit-regularized Hilbert space, the bosonic field operators 
are also regularized to new ones which we can denote as $\phi^Q_\bfr$ and $\pi^Q_\bfr$ where $Q$ represents some regularization scheme.
These qubit-regularized fields generate a unique algebraic structure associated with each qubit regularization scheme, which is referred to as the {\em qubit embedding algebra} (QEA). 
In particular, we are interested in the regularization schemes that preserve the symmetry of the original theory. Therefore the part of the algebra that guarantees this symmetry cannot change. However, the part of the algebra that is not related to the symmetry is usually not compatible with finite dimensional Hilbert space, and therefore are usually changed. The idea of QEA was originally developed in the D-theory approach where simple QEAs for various quantum field theories were constructed by hand \cite{Brower:2003vy}. In this paper, we develop a systematic way to derive the QEAs starting from the Hilbert space of the traditional theory, which can be viewed as a direct sum of many irreducible representations (irreps) of the symmetry group \cite{Byrnes:2005qx,Tagliacozzo:2014bta,Zohar:2014qma,Klco:2019evd}. We define each qubit regularization scheme as a projection operator $P_Q$ which projects the original infinite dimensional space to some finite set of irreps $Q$. This projection naturally preserves important algebra related to the symmetries of the original theory. Furthermore, it also naturally defines a QEA that depends on $Q$. In this work we explicitly construct simple QEAs for $\O(N)$ lattice spin models and $\SU(N)$ lattice gauge theories. Some of these are the same as the ones discovered earlier in the D-theory context, but we also find new ones. 

Qubit regularization is only the first step in the process of studying a quantum field theory using a quantum computer. The second step is the construction of a convenient qubit Hamiltonian that contains the correct quantum critical point where continuum quantum field theory emerges. One must in principle consider both steps together to find the best approach for a particular problem \cite{Zache:2021lrh}. In this work we only focus on the first step, to learn about the various types of qubit regularizations that are possible. 
Ultimately we would like to identify the simplest qubit regularization scheme and the qubit Hamiltonian that contains the desired quantum critical point.

Our paper is organized as follows. In \cref{sec:nlsm} we develop a systematic approach to qubit-regularize the $\O(N)$ non-linear sigma model and show how the QEAs arises. We fully characterize the QEAs of the $\O(2)$ model in \cref{sec:nlsm-O2}, and calculate several simple examples of QEAs for the $\O(3)$ model in \cref{sec:nlsm-O3}. In \cref{sec:nlsm-ON} we extend the analysis of the $\O(3)$ case to $\O(N)$ and show the existence of a simple QEA for any $N$. In \cref{sec:gauge} we show how to extend our systematic approach to qubit-regularize $\SU(N)$ gauge theories. 
We derive the QEA in the simplest regularization schemes for the $\SU(2)$ lattice gauge theory in \cref{sec:gauge-SU2}, for the $\SU(3)$ lattice gauge theory in \cref{sec:gauge-SU3}, and generalize these analysis to the $\SU(N)$ lattice gauge theory and discuss the physical meaning of each element in the QEA in \cref{sec:gauge-SUN}. Some other regularization schemes for $\SU(3)$ obtained with a mathematical software package called \texttt{GAP} \cite{GAP4} are also discussed in \cref{sec:gauge-SU3}. Finally in \cref{sec:discussion}, we discuss the nice features of our simple regularization scheme in $\SU(N)$ gauge theories and present some conjectures about QEAs that could emerge for any $N$.

\section{Lattice spin models}\label{sec:nlsm}
In order to understand the algebraic structure that qubit regularization imposes on the Hilbert space of lattice field theories, it is instructive to begin with lattice spin models. Let us consider $\O(N)$ invariant non-linear sigma models in the Hamiltonian formalism on a lattice. These can be described through $\O(N)$ quantum rotor models whose Hamiltonian is given by
\begin{equation}
  H \ =\ \frac{1}{2\beta} \sum_{\bfr,a} L^a_\bfr L^a_\bfr \ -\ \beta \sum_{\<\bfr,\bfr'\>,i} \phi^i_\bfr \phi^i_{\bfr'},
  \label{eq:ON-Hamiltonian}
\end{equation}
where $\bfr$ labels the spatial lattice sites and $\<\cdot, \cdot\>$ refers to nearest neighbors. Here $L^a_\bfr$, $a=1,2, \cdots ,N(N-1)/2$ are the generators of the $\O(N)$ rotations, while $\phi^i_\bfr$, $i=1,2, \cdots, N$ transform in the fundamental representation of $\O(N)$, i.e.,
\begin{equation}
  [L^a_\bfr, \ L^b_{\bfr'}]\ =\ \i f^{abc} L^c_\bfr \delta_{\bfr,\bfr'},
  \quad [L^a_\bfr, \ \phi^i_{\bfr'}] \ =\ \i (T^a)^i_j \ \phi^j_\bfr\ \delta_{\bfr,\bfr'},
  \label{eq:symalg}
\end{equation}
where $T^a$ is the fundamental representation of $L^a$, and $f^{abc}$ are structure constants of the $\O(N)$ algebra. Since \cref{eq:symalg} is dictated by the symmetry of the theory and specifies how the fields transform under the symmetry, we will refer an algebra of this kind as an \emph{extended symmetry algebra}.

In the traditional lattice non-linear sigma models, $\phi^i_\bfr$ are usually the position operators on a certain homogeneous target space associated with a site $\bfr$, $L^a_\bfr$ generate translations on this target space and therefore can be viewed as momentum operators\footnote{More precisely, when projected to the tangent spaces of the target space, $L^a_\bfr$ give the momentum operators.}, and the local Hilbert space is the space of square integrable functions on this target space. In our case, the target space is an $N-1$ dimensional sphere $S^{N-1} \cong \O(N)/\O(N-1)$, and the Hilbert space $\mathcal{H}$ is $L^2(S^{N-1})$. Notice that the position eigenstates simultaneously diagonalize the $\phi^i_\bfr$ operators, and we have the following relations,
\begin{equation}
  [\phi^i_\bfr, \ \phi^j_{\bfr'}] = 0,\quad \sum_i \ \phi^i_\bfr \phi^i_\bfr = 1.
  \label{eq:extra}
\end{equation}

Since the goal here is to recover $\O(N)$ invariant QFTs in the continuum limit, the exact form of the Hamiltonian in \cref{eq:ON-Hamiltonian} is not important as long as it can be tuned to the correct quantum critical point where the QFT emerges. In order to accomplish this it is important to preserve the symmetries of the theory, which in our case is the $\O(N)$ symmetry. Therefore it is important to preserve the extended symmetry algebras \cref{eq:symalg}. On the other hand, the relations in \cref{eq:extra} are not necessary from the perspective of symmetry. It turns out that \cref{eq:symalg,eq:extra} can only be realized via an infinite dimensional local Hilbert space. Therefore \cref{eq:extra} has to be sacrificed in order to have a finite dimensional local Hilbert space. Since QEA is a local concept, it is independent of the site $\bfr$. For this reason, in our discussions below we will suppress this spatial index $\bfr$ in the operators $L^a_\bfr$ and $\phi^i_\bfr$ and simply refer to them as $L^a$ and $\phi^i$.

One systematic way to ``qubit-regularize'' the Hilbert space to a finite dimensional one, while preserving \cref{eq:symalg} is through working with the ``momentum eigen-blocks'' of the Hilbert space $L^2(S^{N-1})$, i.e., decomposing $L^2(S^{N-1})$ into irreps of the symmetry algebra $\O(N)$. From the standard results of harmonic analysis on spheres, we know that
\begin{align}\label{eq:L2-SN-1}
  L^2(S^{N-1}) = \bigoplus_{l=0}^\infty V_l,
\end{align}
where $V_l$ is the space of spherical harmonics of degree $l$. Furthermore, each $V_l$ forms an irrep of $\O(N)$ with $\dim V_l = \binom{N+l-1}{l} - \binom{N+l-3}{l-2}$. Then we can choose a finite set of integers denoted by $Q$, and truncate the Hilbert space to
\begin{align}\label{eq:ON-HQ}
  \mathcal{H}_Q := \bigoplus_{l\in Q} V_l.
\end{align}
We can define a projector to this space by
\begin{align}
  P_Q := \sum_{l\in Q} P_l,
\end{align}
where $P_l$ is a projector to $V_l$. Then the operators in the truncated theory is simply defined to be the original operators projected to $\mathcal{H}_Q$, i.e., $L^a_Q := P_QL^a P_Q$ and $\phi^i_Q := P_Q \phi^i P_Q$. Since $[L^a, P_Q] = 0$, it is easy to verify that
\begin{align}
  [L^a_Q, L^b_Q] = \i f^{abc} L^c_Q,\qquad [L^a_Q, \phi^i_Q] = \i (T^a)^i_j \phi^j_Q,
\end{align}
i.e., the extended symmetry algebra \cref{eq:symalg} is preserved. Since $\phi^i$ are raising and lowering operators between different $l$'s, the commutator $[\phi^i_Q, \phi^j_Q] \neq 0$. We will show that $L^a_Q$, $\phi^i_Q$ generate a QEA that depends on the choice of $Q$. To illustrate this idea more concretely we will work out explicit examples in some simple cases below and then extend the idea to lattice gauge theories in the next section.

\subsection{The \texorpdfstring{$\O(2)$}{O(2)} lattice spin model}\label{sec:nlsm-O2}

We begin with the simplest example of the $\O(2)$ spin model on the lattice which has been considered earlier in \cite{PhysRevB.103.245137,Meurice:2019ddf}, but as far as we can tell the idea of QEA was not fully explored. The traditional lattice theory is constructed using three operators, the angular momentum operator $L$ that generates $\O(2)$ tranformations and the fields $\phi^1$ and $\phi^2$. The extended symmetry algebra that replaces \cref{eq:symalg} is now given by
\begin{align}
  [L, \phi^\pm] = \pm \phi^\pm,
  \label{eq:symalg-o2}
\end{align}
where we define $\phi^\pm = \frac{1}{\sqrt{2}}(\phi^1 \pm \i \phi^2)$. In the non-linear realization of the traditional lattice theory, we further impose the relations
\begin{align}
  [\phi^1,\phi^2] = 0, \quad (\phi^1)^2 + (\phi^2)^2  = 1,
  \label{eq:extra-o2}
\end{align}
which are not required from the symmetry perspective. These extra relations force the Hilbert space at each lattice site to be infinite dimensional. 

The idea of qubit regularization of the $\O(2)$ model is to define three new operators $L_Q$, $\phi^1_Q$ and $\phi^2_Q$, which act on a finite dimensional local Hilbert space, satisfy the extended symmetry algebra \cref{eq:symalg-o2}, but replace the extra relations \cref{eq:extra-o2} with something else. To accomplish this systematically we first construct the orthonormal ``position'' basis that are eigenstates of the field operators $\phi^1$ and $\phi^2$. These are given by $|\theta\>$, $0\leq \theta < 2\pi$ and satisfy
\begin{align}
  \< \theta|\theta'\> = \delta(\theta-\theta'),\qquad \int_0^{2\pi} \d\theta \ |\theta\>\< \theta|\ =\ 1.
\end{align}
In this basis $\phi^1$ and $\phi^2$ are diagonal, $\phi^1 |\theta \> = \cos\theta|\theta\>$, $\phi^2 |\theta\> = \sin\theta|\theta\>$, but the ``angular momentum'' operator $L$ is not. In order to construct the qubit regularized fields $L_Q$, $\phi^1_Q$ and $\phi^2_Q$, we have to work in the basis that naturally expresses the Hilbert space as a direct sum of irreps of the $\O(2)$ symmetry. These are eigenstates of the angular momentum operator, given by $|k\>, k = 0,\pm 1,\pm 2,\cdots$. They are related to the position eigenstates $|\theta\>$ through the relation
\begin{align}
  \< \theta|k \>\ =\  \frac{1}{\sqrt{2\pi}} \e^{\i k \theta},
\end{align}
i.e., $|k\>$ is related to $|\theta\>$ via Fourier transform. As we will see later, in all the $\O(N)$ models and $\SU(N)$ gauge theories we are going to discuss, the relations between ``position'' eigenstates and ``momentum'' eigen-blocks can be viewed as certain generalizations of the Fourier transform. While $L$ is diagonal in the angular momentum basis $L|k\> = k|k\>$, the field operators $\phi^1$ and $\phi^2$ are not. In fact $\phi^\pm$ act as raising and lowering operators of angular momenta
\begin{align}
  \phi^\pm |k\> = \frac{1}{\sqrt{2}}|k\pm 1\>,
  \label{eq:cranrel-o2}
\end{align}
for all values of $k$. 

We can now qubit-regularize our theory by projecting the traditional Hilbert space to a finite dimensional subspace using the projector
\begin{align}
  P_Q \ =\ \sum_{k \in Q} |k \>\< k |,
\end{align}
where $Q$ is a set of allowed irreps of the $\O(2)$ symmetry. We can then define $L_Q = P_Q L P_Q$, $\phi^\pm_Q = P_Q \phi^\pm P_Q$ as the qubit regularized fields. One simple choice is $Q\ =\ \{k \ | \ k_{\min} \leq k \leq k_{\max}\}$, which is a single block of consecutive integers. This implies $\phi^+_Q |k_{\max}\> = 0$ and $\phi^-_Q |k_{\min}\> = 0$. In this regularization scheme the original infinite dimensional representation is replaced by a $d = k_{\max} - k_{\min} +1$ dimensional Hilbert space. It is easy to verify that the operators $L_Q$ and $\phi^\pm_Q$ are now $d \times d$ matrices that satisfy the extended symmetry algebra \cref{eq:symalg-o2}, but not \cref{eq:extra-o2}. We can construct the operators explicitly by computing the matrix elements:
\begin{align}
  (L_Q)_{k k'} \ =\ \< k | L |k'\> \ =\  k\ \delta_{k,k'},\qquad 
  (\phi^\pm_Q)_{k k'} \ =\ \< k | \phi^\pm |k'\>\ = \ \frac{1}{\sqrt{2}}\delta_{k,{k'\pm 1}}
\end{align}
where $k_{\min} \leq k, k' \leq k_{\max}$. Starting from this matrix representation of $L_Q$ and $\phi^\pm_Q$, we can construct linear combinations of nested commutators until no new matrices are generated. They form a closed Lie algebra, which is what we called the QEA.\footnote{In the definition of an algebra of commutators one usually chooses to work with matrices which are traceless. In our case the only matrix that may not be traceless is $L_Q$. We can always shift $L_Q$ by a proper multiple of identity to make it traceless.} As we prove in \cref{app:O2}, the QEA for $Q\ =\ \{k \ | \ k_{\min} \leq k \leq k_{\max}\}$ is the Lie algebra of the $\SO(d)$ or $\Sp(d/2)$ when $d$ is odd or even respectively.


What is the QEA if we choose $Q$ to be a set of consecutive integer blocks that are themselves not consecutive, for example, $Q = \{1, 2, 3, 5, 6, 7, 8 \}$? This question is perhaps not so interesting since it is unlikely one will choose such a complicated qubit regularization scheme, but it is a valid mathematical question. Since $\phi^\pm$ only connect consecutive integer blocks, $L_Q$ and $\phi^\pm_Q$ are block diagonal. We already know when projected to each consecutive integer block in $Q$, the QEA is either $\SO(d)$ or $\Sp(d/2)$, but the full QEA is not necessarily a direct product of $\SO(d)$ and $\Sp(d/2)$, because the components can be possibly correlated. From Goursat's lemma \cite{lang2005algebra}, we know that two simple Lie groups can not be correlated unless they share the same Lie algebra. In our case, both $\SO(d)$ for $d$ odd and $\Sp(d/2)$ for $d$ even are simple Lie groups, and the Lie groups with different $d$'s have different Lie algebras, except for the accidental isomorphisms $\so(3) \cong \sp(1)$ and $\so(5) \cong \sp(2)$, whose possible correlations can be easily ruled out numerically using a mathematical package called \texttt{GAP} \cite{GAP4}. Therefore we conclude that any two blocks cannot be correlated when they have different dimensions $d$. What happens when two blocks have the same dimension $d$? Actually, in this case the two blocks are fully correlated, because up to a shift by identity, the generators $L$ and $\phi^\pm$ are always identical in the two sectors, and therefore the commutators are also always identical. Therefore we know two blocks are independent if they have different dimensions, and they are fully correlated otherwise. However, can there be any correlation among multiple blocks, even though they are pairwise independent? This possibility can be ruled out using Serre's Lemma \cite{ribet1974adic}, which states that pairwise independence implies full independence for perfect groups, which include $\SO(d)$ and $\Sp(d/2)$. In summary, if the distinct dimensions of the consecutive integer blocks in $Q$ are $d_1, d_2, \cdots, d_m$, where ``distinct'' means that if two $d$'s are equal, we only count them once, then the QEA is $\SO(d_1) \times \SO(d_2) \times \cdots \times \SO(d_m)$ assuming all $d_i$'s are odd. If any one of the $d_i$ is even, we just replace the corresponding block with $\Sp(d_i/2)$. In the example $Q = \{1, 2, 3, 5, 6, 7, 8 \}$, the QEA would be $\SO(3) \times \Sp(2)$. From this example we see clearly that the QEA will depend on the choice of the regularization scheme $Q$.

\subsection{The \texorpdfstring{$\O(3)$}{O(3)} lattice spin model}\label{sec:nlsm-O3}
Let us extend the above analysis to the traditional lattice $\O(3)$ spin model and explore the corresponding QEAs. In this case the lattice model is constructed with quantum fields $L^a, a=1,2,3$ which are the generators of the $\O(3)$ group, and $\phi^i, i=1,2,3$ which transform as a vector under $\O(3)$. The extended symmetry algebra is now given by 
\begin{align}
  [L^a, L^b] \ =\ \i\varepsilon^{abc}L^c,  \quad [L^a, \phi^i] \ =\ \i\varepsilon^{aij} \phi^j. 
  \label{eq:symalg-o3}
\end{align}
In the traditional lattice model, we impose the extra constraints
\begin{align}
  [\phi^i, \phi^j] = 0, \quad \sum_i \ (\phi^i)^2 \ =\ 1,
  \label{eq:extra-o3}
\end{align}
which are not necessary from the symmetry perspective. The local Hilbert space that realizes these relations is the space of all square integrable functions on a sphere, denoted by $L^2(S^2)$, which is infinite dimensional. 

In order to qubit-regularize the $\O(3)$ model we will need to construct operators $L^a_Q$ and $\phi^i_Q$ that act on a finite dimensional local Hilbert space, satisfying the extended symmetry algebra \cref{eq:symalg-o3}. To accomplish this, we first construct the orthonormal ``position'' basis that are eigenstates of $\phi^i$. These are given by $|\theta,\varphi\>$, $0\leq \theta\leq \pi$, $0\leq \varphi < 2\pi$, where $\theta$ and $\varphi$ are the spherical coordinates. The position basis states satisfy
\begin{align}
  \< \theta, \varphi|\theta',\varphi'\> \ =\ \frac{1}{\sin\theta}\delta(\theta-\theta')\delta(\varphi-\varphi'), \quad \int \d\Omega \ |\theta,\varphi\>\<\theta,\varphi|\ =\ 1.
\end{align}
In this basis $\phi^i$ are diagonal,
\begin{align}\label{eq:phi-position}
  \phi^1|\theta,\varphi\> = \sin\theta \cos\varphi|\theta, \varphi\>, \quad \phi^2|\theta,\varphi\> = \sin\theta \sin\varphi|\theta,\varphi\>, \quad \phi^3|\theta,\varphi\> = \cos\theta|\theta,\varphi\>,
\end{align}
but not the ``angular momentum'' operators $L^a$. In order to construct $L^a_Q$ and $\phi^i_Q$ we work in the basis that expresses the original Hilbert space as a direct sum of irreps of the $\O(3)$ symmetry group. These are the angular momentum eigenstates labeled by $|\ell, m \>$, $\ell = 0,1,2,\cdots$ and $-\ell \leq m \leq \ell$. They form a complete orthonormal basis,
\begin{align}
  \< \ell,m|\ell',m'\> = \delta_{\ell \ell'} \delta_{m m'}, \quad \sum_{\ell = 0}^\infty \sum_{m = -\ell}^\ell|\ell, m \> \<\ell, m | = 1.
\end{align}
and can be related to the position basis through the relation
\begin{align}
  \<\theta, \varphi|\ell,m\> = Y_\ell^m(\theta,\varphi),
\end{align}
where $Y_\ell^m(\theta,\varphi)$ are the usual spherical harmonics. While $L^a$ are block diagonal in the angular momentum basis, $\phi^i$ are vector operators that mix $\ell$ with $\ell \pm 1$. To understand how, it is convenient to combine $\phi^1$ and $\phi^2$ into raising and lowering operators $\phi^\pm = \frac{1}{\sqrt{2}} (\phi^1 \pm \i \phi^2)$. 
Using the Wigner-Eckart theorem we can show that
\begin{align}
  \mp\phi^\pm |\ell,m\> &= r_{\ell+1, \ell} c_1^\pm|\ell+1,m\pm 1\> + r_{\ell, \ell} c_0^\pm|\ell,m\pm 1\> + r_{\ell-1, \ell} c_{-1}^\pm|\ell-1,m\pm 1\>, \nonumber\\
  \phi^3 |\ell,m\> &= r_{\ell+1, \ell} c_1^3|\ell+1,m\> + r_{\ell, \ell} c_0^3|\ell,m\> + r_{\ell-1, \ell} c_{-1}^3|\ell-1,m\>,
              \label{eq:field-rel}
\end{align}
where $r_{\ell, \ell'}$ are the reduced matrix elements given by
\begin{align}\label{eq:rme}
  r_{\ell+1,\ell} &= \sqrt{\frac{\ell+1}{2\ell+3}}, \quad r_{\ell,\ell} = 0, \quad r_{\ell-1,\ell} = -\sqrt{\frac{\ell}{2\ell-1}},
\end{align}
and $c_q^{\pm},c^3_q$, $q = 0, \pm 1$ are the Clebsch-Gordan coefficients,
\begin{align}
  c_1^\pm &= \sqrt{\frac{(\ell\pm m+1)(\ell\pm m+2)}{(2\ell+1)(2\ell+2)}}, \quad c_1^3 = \sqrt{\frac{(\ell-m+1)(\ell+m+1)}{(2\ell+1)(\ell+1)}}, \nonumber\\
  c_0^\pm &= \mp \sqrt{\frac{(\ell\mp m)(\ell\pm m+1)}{2\ell(\ell+1)}}, \quad c_0^3 = \frac{m}{\sqrt{\ell(\ell+1)}}, \nonumber\\
  c_{-1}^\pm &= \sqrt{\frac{(\ell\mp m-1)(\ell\mp m)}{2\ell(2\ell+1)}},
             \quad \quad c_{-1}^3 = -\sqrt{\frac{(\ell-m)(\ell+m)}{(2\ell+1)\ell}}.
\end{align}
Using \cref{eq:field-rel} we can compute all matrix elements of $\phi^\pm$ and $\phi^3$ between angular momentum basis states.

In analogy with the $\O(2)$ case we define $P_Q$ as a projector into a subspace of allowed values of $\ell$'s,
\begin{equation}
  P_Q = \sum_{\ell \in Q} \sum_{m = -\ell}^\ell |\ell, m \>\< \ell, m|.
\end{equation}
Using $P_Q$ we can define the qubit regularized fields $L^a_Q = P_Q L^a P_Q$ and $\phi^i_Q = P_Q \phi^i P_Q$. These fields satisfy the extended symmetry algebra \cref{eq:symalg-o3}, but not \cref{eq:extra-o3}. We can construct them explicitly as matrices once we know $\< \ell', m'|L^a|\ell,m\>$ and $\< \ell', m'|\phi^i|\ell,m\>$. The former is simply the spin-$\ell$ representation of angular momentum operators, and we just learned how to compute the latter. Let us consider two simple examples below.

As a first example, we choose $Q = \{0, 1\}$. Then in the basis $(|1,1\>, |1,0\>, |1,-1\>, |0,0\>)^T$ we have
{\setlength\arraycolsep{3pt}}
\begin{align}\label{eq:O3-Lmatrices}
  L^1_Q = \frac{1}{\sqrt{2}}
  \begin{pmatrix}
    0 & 1 & 0 & 0 \\
    1 & 0 & 1 & 0 \\
    0 & 1 & 0 & 0 \\
    0 & 0 & 0 & 0 \\
  \end{pmatrix}, \quad
  L^2_Q = \frac{1}{\sqrt{2}}
  \begin{pmatrix}
    0 & -\i & 0 & 0 \\
    \i & 0 & -\i & 0 \\
    0 & \i & 0 & 0 \\
    0 & 0 & 0 & 0 \\
  \end{pmatrix}, \quad
  L^3_Q =
  \begin{pmatrix}
    1 & 0 & 0 & 0 \\
    0 & 0 & 0 & 0 \\
    0 & 0 & -1 & 0 \\
    0 & 0 & 0 & 0 \\
  \end{pmatrix}.
\end{align}
and
{\setlength\arraycolsep{3pt}}
\begin{align}\label{eq:O3-phimatrices}
  \phi^1_Q = \frac{1}{\sqrt{6}}
  \begin{pmatrix}
    0 & 0 & 0 & -1 \\
    0 & 0 & 0 & 0 \\
    0 & 0 & 0 & 1 \\
    -1 & 0 & 1 & 0 \\
  \end{pmatrix}, \quad
  \phi^2_Q = \frac{1}{\sqrt{6}}
  \begin{pmatrix}
    0 & 0 & 0 & \i \\
    0 & 0 & 0 & 0 \\
    0 & 0 & 0 & \i \\
    -\i & 0 & -\i & 0 \\
  \end{pmatrix}, \quad
  \phi^3_Q = \frac{1}{\sqrt{3}}
  \begin{pmatrix}
    0 & 0 & 0 & 0 \\
    0 & 0 & 0 & 1 \\
    0 & 0 & 0 & 0 \\
    0 & 1 & 0 & 0 \\
  \end{pmatrix}.
\end{align}
The commutators between the six operators $L^a_Q$ and $\phi^i_Q$ can be calculated explicitly,
\begin{align}
  [L^a_Q, L^b_Q] \ =\ \i\varepsilon^{abc}L^c_Q,  \quad [L^a_Q, \phi^i_Q] \ =\ \i\varepsilon^{aij} \phi^j_Q, \quad [\phi^i_Q, \phi^j_Q] = \frac{1}{3} \i \varepsilon^{ija}L^a_Q, 
  \label{eq:O3-0+1}
\end{align}
which is a closed Lie algebra. We can define $J^a_\pm = L^a_Q \pm \sqrt{3}\phi^a_Q$, and then $J^a_+$ and $J^a_-$ form the Lie algebra of two commuting $\SU(2)$ which is isomorphic to the Lie algebra of $\SO(4)$. Thus, we learn that the QEA is the Lie algebra of $\SO(4)$ when $Q = \{0, 1\}$.

A simpler qubit regularization scheme to choose would have been $Q = \{\ell\}$, i.e., a single irrep of $\O(3)$. From \cref{eq:rme} we notice that $r_{\ell,\ell} = 0$, which means $\phi^i_Q = 0$ in this case. Then the QEA is simply the Lie algebra of $\SO(3)$. Qubit models constructed within this regularization scheme will simply be quantum spin-$\ell$ models. Some may find it disturbing that $\phi^i$ has disappeared in this approach and so may feel that we have changed the physics in some fundamental way through this regularization scheme. However, it is important to remember that we are only discussing the Hilbert space in the ultraviolet. The full Hilbert of the lattice theory is much richer and by constructing an appropriate qubit Hamiltonian we can recover all the relevant fields of the original $\O(3)$ quantum field theory at long distances. In fact we can recover the physics of the Wilson-Fisher fixed point at the order-disorder quantum phase transition using an appropriate quantum spin-$\ell$ Hamiltonian \cite{PhysRevLett.72.2777}. 

\subsection{The \texorpdfstring{$\O(N)$}{O(N)} lattice spin model}\label{sec:nlsm-ON}
We can extend the analysis of the previous two subsections to general $\O(N)$ spin models, although a complete discussion can become progressively more complex. In this section we prove a simple result regarding the QEA that one obtains if we choose $Q = \{0, 1\}$ in \cref{eq:ON-HQ}, i.e. the trivial and fundamental representation. From the previous two subsections, we observe that when $N = 2,3$, for this choice of $Q$ the QEA is $\SO(N+1)$. This pattern is actually true for all $N \geq 2$. Let us now give a brief argument for this result.

The traditional Hilbert space of $\O(N)$ models is given by square integrable functions on the $(N-1)$-sphere $L^2(S^{N-1})$. The orthonormal ``position'' basis in this case is labeled by $N-1$ angles $|\theta_1, \theta_2, \cdots, \theta_{N-1}\>$ where $0\leq \theta_i \leq \pi$ for $1 \leq i \leq N-2$, $0\leq \theta_{N-1} < 2\pi$. They are eigenstates of $\phi^i$, with eigenvalues 
\begin{align}
  \phi^1|\theta_1, \theta_2, \cdots, \theta_{N-1}\> &= \cos\theta_1|\theta_1, \theta_2, \cdots, \theta_{N-1}\> \nonumber\\
  \phi^2|\theta_1, \theta_2, \cdots, \theta_{N-1}\> &= \sin\theta_1\cos\theta_2|\theta_1, \theta_2, \cdots, \theta_{N-1}\> \nonumber\\
                          &\enskip \vdots \nonumber\\
  \phi^{N-1}|\theta_1, \theta_2, \cdots, \theta_{N-1}\> &= \sin\theta_1 \cdots \sin\theta_{N-2}\cos\theta_{N-1}|\theta_1, \theta_2, \cdots, \theta_{N-1}\> \nonumber\\
  \phi^N|\theta_1, \theta_2, \cdots, \theta_{N-1}\> &= \sin\theta_1 \cdots \sin\theta_{N-2}\sin\theta_{N-1}|\theta_1, \theta_2, \cdots, \theta_{N-1}\>.
\end{align}
Generalizing the concept of the solid angle we have the following uniform measure on $S^{N-1}$ parameterized by the $N-1$ angles,
\begin{align}
  \d\Omega_{N-1} \ =\  \d\theta_1\d\theta_2 \cdots \d\theta_{N-1} \sin^{N-2}\theta_1 \sin^{N-3}\theta_2 \cdots \sin\theta_{N-2}
\end{align}
such that 
\begin{align}
  \<\theta_1, \cdots, \theta_{N-1}&|\theta'_1, \cdots, \theta'_{N-1}\> \ =\ \frac{1}{\sin^{N-2}\theta_1 \cdots \sin\theta_{N-2}} \delta(\theta_1- \theta_1') \cdots \delta(\theta_{N-1}- \theta_{N-1}'), \nonumber \\
                  &\int \d\Omega_{N-1}\ 
                    |\theta_1, \theta_2, \cdots, \theta_{N-1}\>
                    \<\theta_1, \theta_2, \cdots, \theta_{N-1}|\ =\ 1.
\end{align}
The surface area of $S^{N-1}$ is given by $\Omega_{N-1}\ =\ \frac{2\pi^{N/2}}{\Gamma(N/2)}$.

As before the ``angular momentum'' basis states are simply the irreps of the $\O(N)$ symmetry, and $L^2(S^{N-1})$ is a direct sum these irreps,
\begin{align}\tag{\ref{eq:L2-SN-1}}
  L^2(S^{N-1}) = \bigoplus_{l=0}^\infty V_l.
\end{align}
We know from the standard results from harmonics analysis on spheres that there is a one-to-one correspondence between $V_l$ and degree $l$ homogeneous polynomials in $\mathbb{R}^N$ which are solutions to the Laplacian operator $\nabla^2$. More concretely $V_0$ is simply the space of constant functions over the sphere spanned by a single state which we label as $|0\>$, while $V_1$ is an $N$ dimensional vector space spanned by the coordinate functions on $S^{N-1}$. An orthonormal basis of $V_1$ can be labeled as $|i\>,i=1,2, \cdots, N$, and are related to the ``position'' eigenstates through the following relations,
\begin{align}\label{eq:ON-basis}
  \<\theta_1,\theta_2,\cdots, \theta_{N-1}|0\> &= \sqrt{1/\Omega_{N-1}} \nonumber\\
  \<\theta_1,\theta_2,\cdots, \theta_{N-1}|1\> &= \sqrt{N/\Omega_{N-1}} \cos\theta_1 \nonumber\\
                         &\enskip \vdots \nonumber\\
  \<\theta_1,\theta_2,\cdots, \theta_{N-1}|N-1\> &= \sqrt{N/\Omega_{N-1}} \sin\theta_1 \cdots \sin\theta_{N-2}\cos\theta_{N-1} \nonumber\\
  \<\theta_1,\theta_2,\cdots, \theta_{N-1}|N\> &= \sqrt{N/\Omega_{N-1}} \sin\theta_1 \cdots \sin\theta_{N-2}\sin\theta_{N-1} .
\end{align}
These are eigenstates of $L^a$, but not $\phi^i$. Using these results we can show that
\begin{align}
  \<m|\phi^i|n\> = \frac{1}{\sqrt{N}}(\delta_{m,0}\delta_{n,i} + \delta_{m,i}\delta_{n,0}).
  \label{eq:field-oN}
\end{align}
In the basis \cref{eq:ON-basis}, we can also compute the matrix elements of $L^a$, and it is convenient to relabel the $\frac{1}{2}N(N-1)$ indices $a$ in terms of indices $ij$ where $1 \leq i < j \leq N$,
\begin{align}
  \<m|L^{ij}|n\> = -\i(\delta_{m,i}\delta_{n,j} - \delta_{m,j}\delta_{n,i}).
  \label{eq:amom-oN}
\end{align}
Using \cref{eq:field-oN,eq:amom-oN} we can compute the QEA,
\begin{align}
  [L^{ij}, L^{kl}] &= \i (\delta_{jl}L^{ik} + \delta_{ik}L^{jl} - \delta_{jk}L^{il} - \delta_{il}L^{jk}), \nonumber\\
  [L^{ij}, \phi^k] &= -\i (\delta_{jk}\phi^i - \delta_{ik}\phi^j), \nonumber\\
  [\phi^i, \phi^j] &= \frac{1}{N} \i L^{ij}.
\end{align}
This is the Lie algebra of $\SO(N+1)$. 
A curious fact of these relations is that in the large $N$ limit, this Lie algebra is the same as that of the traditional theory, which can also be understood through the Wigner-In\"on\"u contraction \cite{Inonu:1953sp}. 

\section{Lattice gauge theories}\label{sec:gauge}
We now turn our attention to QEAs that arise in lattice gauge theories. Consider a theory with a gauge group $G$, described by the Kogut-Susskind Hamiltonian \cite{Kogut:1974ag},
\begin{align}
  H = \frac{N}{\beta} \sum_{\<\bfr,\bfr'\>} \big( L_{\bfr\bfr'}^{a2} + R_{\bfr\bfr'}^{a2} \big) - \frac{\beta}{2N} \sum_{\square} \big( U_\square + U_\square^\dagger \big)
  \label{eq:ks}
\end{align}
where the summation is over links $\<\bfr,\bfr'\>$ and plaquettes $\square$. Unlike the spin models, now the quantum degrees of freedom live on links and consist of the electric fields $L^a$, $R^a$, $a = 1, 2, \cdots, \dim G$, and fundamental link operators $U_{ij}, i,j = 1,2,\cdots,N$, where $\dim G$ is the dimension of the group $G$, and $N$ is the dimension of the fundamental representation of $G$. As before we have suppressed the the position indices since they do not play a role in the qubit regularization procedure or in the QEA that emerges. 

In order to preserve the gauge symmetry, it is important that these gauge field operators satisfy the following extended symmetry algebra:
\begin{align}
  [L^a, L^b] &= \i f^{abc} L^c, \quad [L^a, U_{ij}] = T^a_{ik} U_{kj}, \nonumber\\
  [R^a, R^b] &= \i f^{abc} R^c, \quad [R^a, U_{ij}] = -U_{ik} T^a_{kj}, \quad [L^a, R^b] = 0.
             \label{eq:symalg-lgt}
\end{align}
where $T^a$ are the generators of $G$ in the fundamental representation. In traditional lattice gauge theories, the Hilbert space again is the space of square integrable functions on the group manifold $G$, denoted as $L^2(G)$. The link operators $U_{ij}$ can be viewed as ``position'' operators on $G$, while $L^a$ and $R^a$ are ``momentum'' operators that generate left and right translation on $G$. Therefore we further have the following relations among the $U_{ij}$ operators,
\begin{align}
  [U_{ij}, U_{kl}] = 0, \quad [U_{ij},U_{kl}^\dagger] = 0, \quad U^\dagger U = 1,
  \label{eq:extra-lgt}
\end{align}
which are analogous to the relations \cref{eq:extra} in the $\O(N)$ models. Furthermore, the traditional Hilbert space naturally implies the relation
\begin{align}\label{eq:L2=R2}
  \sum_a (L^a)^2 = \sum_a (R^a)^2,
\end{align}
because both of them are the Casimir operators labeling the irreps in \cref{eq:hs-lgt}. The extra relations in \cref{eq:extra-lgt,eq:L2=R2} arise from the choice of the Hilbert space and are not related to the gauge symmetry of the model. In addition to these relations the physical Hilbert space of the problem is much smaller and is obtained by imposing the Gauss' law. We will not worry about this issue here, because we are only focusing on the local Hilbert space structure, and Gauss' law can be imposed after qubit regularizing the Hilbert space.

Qubit regularization of gauge theories consists of defining field operators $L^a_Q$, $R^a_Q$ and $U^Q_{ij}$ that act on a finite dimensional Hilbert space but preserve the gauge invariance by satisfying the extended symmetry algebra \cref{eq:symalg-lgt}. The additional relations \cref{eq:extra-lgt,eq:L2=R2} of the traditional models can be sacrificed if necessary. To accomplish this, let us first understand the Hilbert space structure of the traditional lattice gauge theory. As in the spin models we can choose a basis of ``position'' eigenstates $|g\>$, where $g\in G$ is an element of the group. They form a complete orthonormal basis of $L^2(G)$, i.e.,
\begin{align}
  \< g|g'\> = \delta(g-g'),\quad 
  \int \d g |g\>\< g| = 1,
\end{align}
where $\d g$ is the Haar measure on the group manifold. In this basis all the link operators $U_{ij}$ are diagonal,
\begin{align}
  U_{ij}|g\> = D^f_{ij}(g)|g\>,
\end{align}
where $D^f(g)$ is an $N \times N$ matrix that corresponds to the fundamental representation of $g$. 

In order to qubit-regularize the theory while preserving gauge invariance, we have to go from the ``position'' eigenstates to ``momentum'' eigenstates, where the infinite dimensional local Hilbert space is decomposed into a direct sum of irreps of the symmetry group. This can be accomplished using the Peter-Weyl theorem, which states that $L^2(G)$ decomposes into irreps of $G_L \times G_R$ as \cite{Harlow:2018tng}
\begin{align}
  L^2(G) = \bigoplus_{\lambda \in \hat G} V_\lambda \otimes V_\lambda^*,
  \label{eq:hs-lgt}
\end{align}
where $\hat G$ denotes the set of irreps of $G$. Furthermore, the Peter-Weyl theorem also tells us that the space labeled the irrep $\lambda$ is spanned by the orthonormal basis $|D^\lambda_{ij}\>$,
\begin{align}
  \<D^\lambda_{ij}|D^{\lambda'}_{kl}\> \ =\ \delta_{\lambda\lambda'} \delta_{ik} \delta_{jl},\qquad \sum_{\lambda\in \hat G} \sum_{i,j} |D^{\lambda}_{ij}\> \<D^{\lambda}_{ij}|\ =\ 1,
\end{align}
and are related to the ``position'' eigenstates by
\begin{align}
  \<g|D^\lambda_{ij}\> = \sqrt{d_\lambda} D^\lambda_{ij}(g),
\end{align}
where $D^\lambda(g)$ is the matrix representation of $g$ in the irrep $\lambda$, and $d_\lambda$ is the dimension of the representation $\lambda$. For example, the state $|0\>$ in the trivial representation satisfies $\<g|0\>=1$. The operators $L^a$ and $R^a$ are block diagonal in this basis with matrix elements
\begin{align}
  \<D^\lambda_{ij}|L^a|D^\lambda_{kl}\> = (T^a_\lambda)_{ik}\delta_{jl},\qquad \<D^\lambda_{ij}|R^a|D^\lambda_{kl}\> = -(T^a_\lambda)_{lj}\delta_{ik},
  \label{eq:LRmat-suN}
\end{align}
where $T^a_\lambda$ are the corresponding generators of $G$ in the representation $\lambda$. On the other hand when $U_{ij}$ acts on $|D^\lambda_{ij}\>$, states in other irreps are generated. This can be understood by noting that
\begin{align}\label{eq:U-action}
  U_{kl} |D^\lambda_{ij}\> &= \int \d g \sqrt{d_\lambda} D^\lambda_{ij}(g) D^f_{kl}(g) |g\>,
\end{align}
where $D^\lambda_{ij}(g)D^f_{kl}(g)$ is an element in the tensor product representation $\lambda \otimes f$. In the case of $G = \SU(N)$, we know that all irreps can be labeled by the Young diagrams. In particular, the fundamental representation $\mathbf{N}$ is labeled by a single box $\tyng{1}$, and $\lambda \otimes f$ decomposes into irreps that correspond to adding a single box to the Young diagram of $\lambda$.

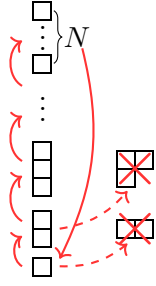
\begin{figure}[htb]
  \centering
  \begin{tikzpicture}[baseline={([yshift=-.5ex]current bounding box.center)}]
    \node at (0,0) [right] {$\tyng{1}$};
    \node at (0,0.5) [right] {$\tyng{1,1}$};
    \node at (0,1.3) [right] {$\tyng{1,1,1}$};
    \node at (0,2.2) [right, xshift = 2.5pt] {$\vdots$};
    \node at (0,2.7) [right] {$\tyng{1}$};
    \node at (0,3.15) [right, xshift = 2.5pt] {$\vdots$};
    \node at (0,3.4) [right] {$\tyng{1}$};
    \draw [decorate,decoration={brace,amplitude=3pt},xshift=12pt,yshift=0pt](0,3.4) -- (0,2.7) node [midway,right] {$N$};
    \path[thick,red!80,->] (0,0) edge [bend left=60] (0,0.4);
    \path[thick,red!80,->] (0,0.6) edge [bend left=60] (0,1.2);
    \path[thick,red!80,->] (0,1.4) edge [bend left=60] (0,2);
    \path[thick,red!80,->] (0,2.4) edge [bend left=60] (0,3);
    \path[thick,red!80,->] (0.8,2.9) edge [bend left=20] (0.5,0.1);
    \node at (1.5,0.5) {$\tyng{2}$};
    \node[red!80] at (1.5,0.5) {\Huge$\times$};
    \path[thick,red!80,dashed,->] (0.5,0) edge [bend right=20] (1.4,0.3);
    \node at (1.5,1.3) {$\tyng{2,1}$};
    \node[red!80] at (1.5,1.3) {\Huge$\times$};
    \path[thick,red!80,dashed,->] (0.5,0.5) edge [bend right=20] (1.4,1);
  \end{tikzpicture}
  \caption{A simple qubit regularizarion scheme for $\SU(N)$ gauge theory by keeping only the anti-symmetric representations. In this case $U$ acts on the irreps by adding a single box to the column, and therefore acts as a cyclic permutations on the anti-symmetric representations.}
  \label{fig:truncation}
\end{figure}

In order to qubit-regularize the theory, we will project the full Hilbert space given in \cref{eq:hs-lgt} to a finite dimensional one that only contains some of the irreps $\lambda$. While there are many choices, let us focus on a simple regularization scheme by choosing only the anti-symmetric representations,
\begin{align}
  Q = \{\mathbf{N}, \bigwedge\nolimits^2 \mathbf{N}, \cdots, \bigwedge\nolimits^N \mathbf{N} \cong \mathbf{1}\},
  \label{eq:qr-suN}
\end{align}
where $ \bigwedge^k \mathbf{N}$ is the irrep corresponding to Young diagrams with $k$ boxes arranged in a single column. All the other representations will be projected out. The action of $U$ on such a regularized Hilbert space has a simple form and is shown diagrammatically in  \cref{fig:truncation}.
As we can see, $U$ acts cyclically in the space of irreps, beacuse the irrep with $N$ boxes in a column is the trivial irrep and adding a box takes to the trivial irrep takes you back to the irrep with a single box. This regularization scheme is particularly simple for $N=2,3$ because all matrix elements of $U$ in the regularized space can be written as certain integrals over $\SU(N)$ which can be evaluated simply using the invariance of the Haar measure, without the need to perform explicit integration over the group. 
For example, in the case of $\SU(2)$ there are only two irreps in this simple regularization scheme, the singlet and the fundamental $\tyng{1}$. Then there are two non-zero matrix elements given by 
\begin{align}\label{eq:umat-su2}
  \< D^f_{ij}| U_{kl} |0\> &= \sqrt{2}\int \d g D^f_{ij}(g)^* D^f_{kl}(g) = \frac{1}{\sqrt{2}} \ \delta_{ik} \delta_{jl}. \nonumber \\
  \< 0| U_{kl} |D^f_{ij}\> &= \sqrt{2}\int \d g D^f_{kl}(g) D^f_{ij}(g) = \frac{1}{\sqrt{2}} \ \varepsilon_{ki} \varepsilon_{lj}.
\end{align}
In the case of $\SU(3)$ there are three irreps, the singlet, the fundamental $\tyng{1}$ and anti-fundamental $\tyng{1,1}$. In this case we need three matrix elements
\begin{align}\label{eq:umat-su3}
  \< D^f_{ij}| U_{kl} |0\> &= \sqrt{3}\int \d g D^f_{ij}(g)^* D^f_{kl}(g) = \frac{1}{\sqrt{3}}\ \delta_{ik} \delta_{jl}, \nonumber \\
  \< D^{\bar{f}}_{ij}| U_{kl} |D^f_{mn}\> &= 3\int \d g D^f_{ij}(g)D^f_{kl}(g) D^f_{mn}(g) = \frac{1}{2} \ \varepsilon_{ikm} \varepsilon_{jln}, \nonumber \\
  \< 0| U_{kl} |D^{\bar{f}}_{ij}\> &= \sqrt{3}\int \d g D^f_{kl}(g) D^f_{ij}(g)^* = \frac{1}{\sqrt{3}} \ \delta_{ki} \delta_{lj}. 
\end{align}
Further details on how to evaluated the above integrals are given in \cref{app:U-matrix-elements}. Armed with these results we derive the QEAs for our simple regularization scheme in the case of $G = \SU(2)$ and $G = \SU(3)$.

\subsection{\texorpdfstring{$\SU(2)$}{SU(2)} lattice gauge theory}\label{sec:gauge-SU2}

Using the fact $L^2(\SU(2)) = L^2(S^3)$, we expect the QEA for $\SU(2)$ lattice gauge theory to be the same as that of the $\O(4)$ spin model when the two subspaces are regularized in the same way.  In particular, when truncated to the trivial and fundamental representations, the QEA is expected to be $\SO(5)$. Let us derive this again here but from the perspective of a lattice gauge theory. The local quantum fields of the traditional model are given by $L^a,R^a, a = 1,2,3$ and $U_{ij},i,j=1,2$. Let $P_Q$ be a projector into a subspace of the irreps 
\begin{equation}
  P_Q = \sum_{\lambda \in Q} \sum_{i,j} |D^\lambda_{ij}\>\< D^\lambda_{ij}|,
\end{equation}
Using $P_Q$ we can define the qubit-regularized fields $L^a_Q$, $R^a_Q$ and $U_{ij}^Q$ as
\begin{equation}
  L^a_Q = P_Q L^a P_Q, \quad R^a_Q = P_Q R^a P_Q, \quad U_{ij}^Q = P_Q U_{ij} P_Q.
\end{equation}
For our simple choice of qubit regularization given in \cref{eq:qr-suN} we get $Q = \{\mathbf{2},\mathbf{1}\}$, i.e. the fundamental representation and the trivial representation. We can compute all the matrix elements of $L^a_Q$, $R^a_Q$ using \cref{eq:LRmat-suN}, and
in the basis $(|D^f_{11}\>, |D^f_{12}\>, |D^f_{21}\>, |D^f_{22}\>, |0\>)^T$, we get the following $5\times 5$ matrices:
{\setlength\arraycolsep{3pt}}
\begin{align}
  L^1_Q &=
  \begin{pmatrix}
    0 & 0 & 1 & 0 & 0 \\
    0 & 0 & 0 & 1 & 0 \\
    1 & 0 & 0 & 0 & 0 \\
    0 & 1 & 0 & 0 & 0 \\
    0 & 0 & 0 & 0 & 0 \\
  \end{pmatrix}, \quad\qquad
  L^2_Q =
  \begin{pmatrix}
    0 & 0 & -\i & 0 & 0 \\
    0 & 0 & 0 & -\i & 0 \\
    \i & 0 & 0 & 0 & 0 \\
    0 & \i & 0 & 0 & 0 \\
    0 & 0 & 0 & 0 & 0 \\
  \end{pmatrix}, \quad
  L^3_Q =
  \begin{pmatrix}
    1 & 0 & 0 & 0 & 0 \\
    0 & 1 & 0 & 0 & 0 \\
    0 & 0 & -1 & 0 & 0 \\
    0 & 0 & 0 & -1 & 0 \\
    0 & 0 & 0 & 0 & 0 \\
  \end{pmatrix}, \\
  R^1_Q &=
  \begin{pmatrix}
    0 & -1 & 0 & 0 & 0 \\
    -1 & 0 & 0 & 0 & 0 \\
    0 & 0 & 0 & -1 & 0 \\
    0 & 0 & -1 & 0 & 0 \\
    0 & 0 & 0 & 0 & 0 
  \end{pmatrix}, ~
  R^2_Q =
  \begin{pmatrix}
    0 & -\i & 0 & 0 & 0 \\
    \i & 0 & 0 & 0 & 0 \\
    0 & 0 & 0 & -\i & 0 \\
    0 & 0 & \i & 0 & 0 \\
    0 & 0 & 0 & 0 & 0 \\
  \end{pmatrix}, ~
  R^3_Q =
  \begin{pmatrix}
    -1 & 0 & 0 & 0 & 0 \\
    0 & 1 & 0 & 0 & 0 \\
    0 & 0 & -1 & 0 & 0 \\
    0 & 0 & 0 & 1 & 0 \\
    0 & 0 & 0 & 0 & 0 \\
  \end{pmatrix}.
\end{align}
Similarly $U^Q_{ij}$ can be computed as $5\times 5$ matrices using \cref{eq:umat-su2},
\begin{align}
  U_{11}^Q &= \frac{1}{\sqrt{2}}
  \begin{pmatrix}
    0 & 0 & 0 & 0 & 1 \\
    0 & 0 & 0 & 0 & 0 \\
    0 & 0 & 0 & 0 & 0 \\
    0 & 0 & 0 & 0 & 0 \\
    0 & 0 & 0 & 1 & 0 \\
  \end{pmatrix}, \quad
  U_{12}^Q = \frac{1}{\sqrt{2}}
  \begin{pmatrix}
    0 & 0 & 0 & 0 & 0 \\
    0 & 0 & 0 & 0 & 1 \\
    0 & 0 & 0 & 0 & 0 \\
    0 & 0 & 0 & 0 & 0 \\
    0 & 0 & -1 & 0 & 0 \\
  \end{pmatrix}, \nonumber\\
  U_{21}^Q &= \frac{1}{\sqrt{2}}
  \begin{pmatrix}
    0 & 0 & 0 & 0 & 0 \\
    0 & 0 & 0 & 0 & 0 \\
    0 & 0 & 0 & 0 & 1 \\
    0 & 0 & 0 & 0 & 0 \\
    0 & -1 & 0 & 0 & 0 \\
  \end{pmatrix}, \quad
  U_{22}^Q = \frac{1}{\sqrt{2}}
  \begin{pmatrix}
    0 & 0 & 0 & 0 & 0 \\
    0 & 0 & 0 & 0 & 0 \\
    0 & 0 & 0 & 0 & 0 \\
    0 & 0 & 0 & 0 & 1 \\
    1 & 0 & 0 & 0 & 0 \\
  \end{pmatrix}.
\end{align}
It can be checked directly that $L^a_Q$, $R^a_Q$ and the independent Hermitian matrices constructed using $U^Q_{ij}$ form $\SO(5)$ algebra, as we already mentioned at the beginning of this section.

\subsection{\texorpdfstring{$\SU(3)$}{SU(3)} lattice gauge theory}\label{sec:gauge-SU3}

We can extend the above calculations to $\SU(3)$ lattice gauge theory, which is interesting from the perspective of QCD and is already being explored on a quantum computer \cite{PhysRevD.103.094501}. In this case $P_Q$ is again of the form
\begin{align}
  P_Q = \sum_{\lambda \in Q} \sum_{i,j} |D^\lambda_{ij}\>\< D^\lambda_{ij}|,
\end{align}
where instead of \cref{eq:qr-suN} we first choose the simpler case $Q = \{\mathbf{1}, \mathbf{3}\}$, i.e., the trivial representation and the fundamental representation. In this case qubit-regularized Hilbert space is ten dimensional with basis states
\begin{align}
  (|D^f_{11}\>, |D^f_{12}\>, |D^f_{13}\>, |D^f_{21}\>, |D^f_{22}\>, |D^f_{23}\>, |D^f_{31}\>, |D^f_{32}\>, |D^f_{33}\>, |0\>)^T.
\end{align}
Again using \cref{eq:LRmat-suN} we can compute all matrix elements of $L^a_Q$, $R^a_Q$ in this ten dimensional space, which can be written compactly as
\begin{align}
  L^a_Q = \lambda_a \otimes \mathbbm{1}_3 \oplus 0, \quad R^a_Q = - \mathbbm{1}_3 \otimes \lambda_a^* \oplus 0,
\end{align}
where $\lambda_a$, $a = 1, \cdots, 8$ are Gell-Mann matrices. Similarly $U^Q_{ij}$ can also be constructed as $10\times 10$ matrices using \cref{eq:umat-su3}.
Using \texttt{GAP} we can show that
$L^a_Q$, $R^a_Q$, $U_{ij}^Q$ and $U_{ij}^{Q\dagger}$ generate the $\SU(10)$ algebra. In other words, starting with these $34$ matrices we generate $65$ more matrices under commutation relations and obtain the $99$ generators of $\SU(10)$. We we will prove this more generally for the $\SU(N)$ lattice gauge theories in the next section. 

We end this section by considering two more regularization schemes. If we choose 
$Q = \{\mathbf{1}, \mathbf{3}, \mathbf{\bar 3}\}$ and input the matrices into \texttt{GAP} we find the QEA to be $\SU(19)$. Instead, if we choose  $Q = \{\mathbf{3}, \mathbf{\bar 3}\}$, \texttt{GAP} tells us that QEA is $\Sp(9)$. A nice feature of the $\SU(19)$ regularization scheme is that $U^Q_{ij}$ is a cyclic raising operator in the representation space: $\mathbf{1} \rightarrow \mathbf{N} \rightarrow \bar{\mathbf{N}} \rightarrow \mathbf{1}$. This could be a desirable property since it is similar to that of the traditional model.

\subsection{\texorpdfstring{$\SU(N)$}{SU(N)} lattice gauge theory}\label{sec:gauge-SUN}
In this section we will show that if we qubit-regularize an $\SU(N)$ gauge theory with $Q = \{\mathbf{1}, \mathbf{N}\}$, the resulting QEA is $\SU(N^2+1)$. We will also provide a physical interpretation of all the elements in the Lie algebra of $\SU(N^2+1)$.

In this case we can label the $N^2+1$ orthonormal basis states of the regularized Hilbert space as $|\alpha\rangle, \alpha = 0,1,2,\cdots,N^2$, such that $|0\>$ denotes the state in the singlet, and $|D^f_{ij}\rangle =: |\alpha = N(i-1)+j\>$. We also define $U_{ij}^Q =: U_\alpha^Q$, where $1 \leq \alpha = N(i-1)+j \leq N^2$. The result in \cref{app:U-matrix-elements} implies that $\langle\gamma|U^Q_\alpha|\beta\> = \frac{1}{\sqrt{N}} \delta_{\gamma\alpha}\delta_{0\beta}$, which gives
\begin{align}
U_\alpha^Q &= \frac{1}{\sqrt{N}}e_{\alpha 0}, \qquad U_\alpha^{Q\dagger} = \frac{1}{\sqrt{N}}e_{0\alpha},
\end{align}
where $e_{\alpha\beta}$ is an $(N^2+1) \times (N^2+1)$ matrix with matrix elements $(e_{\alpha\beta})_{\gamma\delta} = \delta_{\alpha\gamma}\delta_{\beta\delta}$. Using this we can show the following commutation relations
\begin{align}
  [U^Q_\alpha, U^Q_\beta] = [U_\alpha^{Q\dagger}, U_\beta^{Q\dagger}] = 0,\quad
  [U_\alpha^Q, U_\beta^{Q\dagger}] = \frac{1}{N} (e_{\alpha \beta} - \delta_{\alpha \beta}e_{00}).
\end{align}
Interestingly, we again observe that this algebra is reduced to the traditional one in \cref{eq:extra-lgt} when $N \rightarrow \infty$. $[U^Q_\alpha, U_\beta^{Q\dagger}]$ generate $N^4$ operators, each of which is an $(N^2+1)\times (N^2+1)$ matrix. The linear combinations of these operators give all traceless matrices within the $N^2\times N^2$ block that does not contain the singlet. More concretely, $[U^Q_\alpha, U_\beta^{Q\dagger}] \propto e_{\alpha \beta}$ for $\alpha \neq \beta$ give all off diagonal elements in this $N^2\times N^2$ block, while $[U^Q_\alpha, U_\alpha^{Q\dagger}] - [U^Q_{\alpha+1}, U_{\alpha+1}^{Q\dagger}] = e_{\alpha \alpha} - e_{(\alpha+1)(\alpha+1)}$ for $\alpha = 1, \cdots, N^2-1$ gives the $N^2-1$ independent traceless diagonal matrices. We denote these $N^4-1$ operators which are represented by traceless matrices within the $N^2\times N^2$ block as $t_{\alpha\beta}$ such that
\begin{align}
t_{\alpha\beta} = 
\begin{cases}
 \e_{\alpha\beta}\ &\mbox{when } \alpha\neq \beta,   \\
 e_{\alpha \alpha} - e_{(\alpha+1)(\alpha+1)} &\mbox{when } \alpha= \beta \neq N^2
\end{cases} 
\end{align}
The Hermitian linear combinations of these matrices form the Lie algebra of $\SU(N^2)$. There is one more independent operator
\begin{align}
  E := \frac{N}{N^2+1}\sum_{\alpha = 1}^{N^2} [U_\alpha, U_\alpha^\dagger] = \frac{1}{N^2+1} - e_{00},
\end{align}
which is normalized such that $U_\alpha^Q$ and $U_\alpha^{Q\dagger}$ have charges $\pm 1$ in \cref{eq:EU}. $E$ is traceless and proportional to the identity in the  $N^2\times N^2$ block. The $2N^2$ operators $U_\alpha$ and $U_i^\dagger$, along with $N^4-1$ operators $t_{\alpha\beta}$ and the operator $E$, form the $N^4+2N^2$ dimensional Lie algebra of $\SU(N^2+1)$.

The operator $E$ is the unique element in the Lie algebra of $\SU(N^2+1)$ that commutes with all $L^a$ and $R^a$, which can be argued as follows. The Hilbert space is a direct sum of the $N^2$ basis states $|D^f_{ij}\>$ and the singlet $|0\rangle$, which are two irreps of $\SU(N)_L \times \SU(N)_R$. Using Schur's lemma, we know that any element that commutes with both $L^a$ and $R^a$ must be proportional to identity in each of the irreps. Since the generators of $\SU(N^2+1)$ are all traceless, there is a unique element that commutes with both $L^a$ and $R^a$, which is nothing but the $E$ operator. Furthermore, it is easy to verify that $E$ also satisfies
\begin{align}\label{eq:EU}
  [E, U_{ij}^Q] = U_{ij}^Q, \quad [E, U_{ij}^{Q\dagger}] = -U_{ij}^{Q\dagger}.
\end{align}
Therefore, the operator $E$ can be viewed as the generator of a $\U(1)$ gauge field, and the link operators $U_{ij}^Q$ and $U_{ij}^{Q\dagger}$ carry opposite charges under $E$. Thus, $Q = \{\mathbf{1},\mathbf{N}\}$ could also be used as a qubit regularization of a $\U(N)$ gauge theory. This is reminiscent of the D-theory approach where $\SU(N)$ was implemented using the QEA of $\SU(2N)$ and again in that regularization there was an element similar to $E$ which could be used to construct a $\U(N)$ gauge theory.

Let us now understand how the $N^4-1$ operators $t_{\alpha\beta}$, that appear in the Lie algebra of $\SU(N^2)$, transform under the $\SU(N)_L \times \SU(N)_R$ subgroup that is embedded in it. In the operators $t_{\alpha\beta}$, the index $\alpha$ transforms as $\mathbf{N}_L \otimes \mathbf{\bar N}_R$, while the index $\beta$ transforms as $\mathbf{\bar{N}}_L \otimes \mathbf{N}_R$. Hence together with $E$, the operators $t_{\alpha\beta}$ transform as
\begin{align}
  \mathbf{N}_L \otimes \mathbf{\bar N}_R \otimes \mathbf{\bar{N}}_L \otimes \mathbf{N}_R &\cong (\mathbf{N} \otimes \mathbf{\bar N})_L \otimes (\mathbf{\bar N} \otimes \mathbf{N})_R \nonumber\\
  &\cong (\adj \oplus \mathbf{1})_L \otimes (\adj \oplus \mathbf{1})_R \nonumber \\
 &\cong \adj_L \otimes \adj_R \oplus \adj_L \oplus \adj_R \oplus \mathbf{1}.
\end{align}
In this decomposition, $\adj_L$ and $\adj_R$ are nothing but $L^a$ and $R^a$, while $\mathbf{1}$ is the $E$ operator. The remaining part $\adj_L \otimes \adj_R$ is exactly the way adjoint link operators $U^{\adj,Q}_{ij}$ would transform. Thus, we also get the adjoint links for free.

In summary, we learn that the qubit regularization with the QEA $\SU(N^2+1)$ naturally contains the electric operators $L^a_Q$, $R^a_Q$ and $E$, the fundamental link operators $U^Q_{ij}$ and $(U^Q_{ij})^\dagger$, and the adjoint link operators $U^{\adj,Q}_{ij}$ that are self-dual. This discussion also motivates to define a qubit regularization of an $\SU(N)$ gauge theory with adjoint link operators using an $N^2$ dimensional Hilbert space, with the QEA of $\SU(N^2)$.

\section{Discussion and Conclusions}\label{sec:discussion}

Using examples from spin models and gauge theories, in this work we showed that qubit regularizations are characterized by an algebraic structure referred to as the qubit emdedding algebra (QEA). We propose to use QEA along to define the qubit regularization scheme. For example, we showed how the traditional $\O(N)$ spin model can be qubit-regularized using the $\SO(N+1)$ scheme. We also showed that the traditional $\SU(2)$ gauge theory can be qubit-regularized using $\SO(5)$ scheme, while the $\SU(N)$ gauge theory can be qubit-regularized using $\SU(N^2+1)$ scheme for $N\geq 3$. For $N = 3$ we discovered that there are also $\Sp(9)$ or $\SU(19)$ schemes, using the mathematical software package \texttt{GAP}. Based on these numerical results using \texttt{GAP}, we conjecture that for $Q = \{\mathbf{1}, \mathbf{N}, \bigwedge^2\mathbf{N},\cdots, \bigwedge^{N-1}\mathbf{N} \}$ the QEA is $\SU\big(\binom{2N}{N} - 1\big)$, while for $Q = \{\mathbf{N},\mathbf{\bar N}, \mathbf{1}\}$ and $N \geq 4$, the QEA is $\SO(2N^2+1)$. The latter conjecture is based on the result from the $\SU(3)$ gauge theory, that if we choose $Q = \{\mathbf{3},\mathbf{\bar 3}, \mathbf{1}\}$ and set the matrix elements $\< D^{\bar f}_{ij}| U_{kl} | D^f_{mn}\>$ to be zero, which vanishes for $N\geq 4$, then the QEA is $\SO(19)$.

The idea of qubit regularization was introduced long ago within the D-theory approach \cite{Chandrasekharan:1996ih,Brower:1997ha,Brower:2003vy}, but not within a systematic approach that we have adopted in our work here. Interestingly, some of the QEAs that were proposed earlier are the same as the ones we found here. For example the $\O(N)$ spin models were also qubit-regularized using the QEA of $\SO(N+1)$ in \cite{Brower:2003vy}. Even the qubit regularization of the $\SU(2)$ gauge theory using the QEA of $\SO(5)$ was known earlier \cite{Chandrasekharan:1996ih}.
The representation proposed in the D-theory approach was a four-dimensional spin representation of $\SO(5)$, i.e., $\Sp(2)$, while the one we naturally found here from the traditional model is the five-dimensional fundamental representation. This is yet another important feature of QEA to keep in mind --- the QEA alone does not fully determine the qubit regularization, we also need to specify the representation of the QEA that is used to construct the quantum fields.

In the D-theory approach, $\SU(N)$ lattice gauge theories for $N\geq 3$, were qubit-regularized using the QEA of $\SU(2N)$ in the fundamental representation. On the other hand in our work we found regularization schemes with QEA of $\SU(N^2+1)$ in the fundamental representation. Interestingly, in these regularization schemes there is always an additional operator $E$ that generates $\U(1)$ gauge transformations, which means the same QEAs can also be used to regularize a $\U(N)$ lattice gauge theory. Moreover, we showed how in our $\SU(N^2+1)$ regularization scheme, the adjoint link operators are also generated naturally, and in a gauge theory with only adjoint link operators, we can simply regularize the theory with in the $\SU(N^2)$ scheme. In this case, no new operators are generated in this scheme.

A significant distinction between the regularization schemes discussed in this paper as compared to earlier schemes \cite{Chandrasekharan:1996ih,Brower:1997ha,Brower:2003vy} is that the relation $\sum_a (L^a)^2 = \sum_a (R^a)^2$ in \cref{eq:L2=R2} is preserved in the former but not in the latter. In traditional lattice gauge theories, this relation arises from the fact that both $(L^a)^2$ and $(R^a)^2$ are Casimir operators which label the representations in \cref{eq:hs-lgt} and they are equal. Our regularization scheme preserves this structure of the irreps but not the quantum link model. Further studies are necessary to understand whether this difference is important to recover the asymptotically free fixed point.


\acknowledgments{We would like to thank T. Bhattacharya, A. Hulse, I. Marvian, M. Nguyen, H. Singh and R. Gupta for discussions. We would also like to thank MathOverflow user Paul Levy\footnote{\url{https://mathoverflow.net/users/26635/paul-levy}}, for sketching out the idea for the proof of \cref{thm:O2}. The material presented here is based on work supported by the U.S. Department of Energy, Office of Science --- High Energy Physics Contract KA2401032 (Triad National Security, LLC Contract Grant No. 89233218CNA000001) to Los Alamos National Laboratory. S.C. is supported by a Duke subcontract of this grant. S.C. and H.L. are also supported for this work in part by the U.S. Department of Energy, Office of Science, Nuclear Physics program under Award No. DE-FG02-05ER41368.}

\appendix
\section{Proof of the QEAs for the \texorpdfstring{$\O(2)$}{O(2)} model}\label{app:O2}
Let's formulate the result as the following theorem,
\begin{Theorem}\label{thm:O2}
  Let $\phi_d^+$ and $\phi_d^-$ be the $d \times d$ upper shift matrix and lower shift matrix respectively, i.e. $(\phi_d^\pm)_{ij} = \delta_{i\pm1,j}$, where $1\leq i, j \leq d$. Let $\mathfrak{g}_d$ be the Lie algebra generated by $\phi^+_d$ and $\phi^-_d$ over $\mathbb{C}$. Let $J_d = (-1)^i\delta_{i,d+1-i}$ be an anti-diagonal bilinear form. Then $\mathfrak{g}_d$ is the Lie algebra of $d \times d$ matrices that preserve the bilinear form $J_d$, i.e.
  \begin{align}\label{eq:O2-g}
    \mathfrak{g} = \{x \in \mathfrak{gl}(d, \mathbb{C}) | x^T J_d + J_d x = 0\}.
  \end{align}
In particular, since $J_d$ is symmetric (anti-symmetric) when $d$ is odd (even), we know $\mathfrak{g}_d \cong \so(d, \mathbb{C})$ ($\mathfrak{g}_d \cong \sp(d, \mathbb{C})$) when $d$ is odd (even). 
\end{Theorem}

\begin{proof}
Let's assume $d$ is odd, and the case of $d$ being even can be proved in a similar way. It is easy to check that $\phi_d^{\pm T} J_d + J_d \phi_d^\pm = 0$. Therefore $\mathfrak{g}_d \subset \so(d, \mathbb{C})$.
  
In order to show $\mathfrak{g}_d \cong \so(d, \mathbb{C})$, we use induction over $d$. When $d = 3$, we have
\begin{align}
  h_3 := [\phi_3^+, \phi_3^-] =
  \begin{pmatrix}
    1 & 0 & 0 \\
    0 & 0 & 0 \\
    0 & 0 & -1
  \end{pmatrix}.
\end{align}
Clearly $\phi_3^+, \phi_3^-, h_3$ are independent of each other, and therefore $\dim_\mathbb{C}(\mathfrak{g}_3) \geq 3$. Since $\dim_\mathbb{C}(\so(3, \mathbb{C})) = 3$, we know that $\mathfrak{g}_3 \cong \so(3, \mathbb{C})$.

Now assuming $\mathfrak{g}_{d-2} \cong \so(d-2, \mathbb{C})$, let's consider the case for $d$. In this case we have
\begin{align}
  h_d := [\phi_d^+, \phi_d^-] =
  \begin{pmatrix}
    1 & \mathbf{0} & 0 \\
    \mathbf{0} & \mathbf{0}_{d-2} & \mathbf{0} \\
    0 & \mathbf{0} & -1
  \end{pmatrix}
\end{align}
and
\begin{align}
  \phi_d^+ + [\phi_d^+, h_d] &= 
  \begin{pmatrix}
    0 & \mathbf{0} & 0 \\
    \mathbf{0} & \phi_{d-2}^+ & \mathbf{0} \\
    0 & \mathbf{0} & 0
  \end{pmatrix}, \\
  \phi_d^- - [\phi_d^-, h_d] &= 
  \begin{pmatrix}
    0 & \mathbf{0} & 0 \\
    \mathbf{0} & \phi_{d-2}^- & \mathbf{0} \\
    0 & \mathbf{0} & 0
  \end{pmatrix}.
\end{align}
Therefore we see that $\mathfrak{g}_{d} \supset \so(d-2, \mathbb{C})$ using the induction hypothesis. We will complete the proof using the root system of $\so(d, \mathbb{C})$. In order to do so, we need to first determine the Cartan subalgebra of $\so(d, \mathbb{C})$, which can be chosen to be
\begin{align}
  h_d^i := e_{ii} - e_{d+1-i, d+1-i},
\end{align}
where $e_{ij}$ is an $d \times d$ matrix with $1$ at the $ij$ position and zero everywhere else. From this definition we can see that $h_d^1 = h_d$. Then for an element $x\in \mathfrak{g}_d$ which simultaneously diagonalize $h_d^i$ in the adjoint representation, i.e. $[h_d^i, x] = \alpha^i x$, the corresponding root is defined to be $\vec\alpha$. It can be checked that $[h_d, \phi_d^+]$ simultaneously diagonalize $h_d^i$ with eigenvalues $\vec\alpha = (1, -1, 0, \cdots, 0)^T$. Together with the simple roots in $\so(d-2, \mathbb{C})$, i.e.
\begin{align*}
  \begin{pmatrix}
    0 \\ 1 \\ -1 \\ 0 \\ \vdots \\ 0
  \end{pmatrix},
  \begin{pmatrix}
    0 \\ 0 \\ 1 \\ -1 \\ \vdots \\ 0
  \end{pmatrix},
  \cdots,
  \begin{pmatrix}
    0 \\ \vdots \\ 0 \\ 0 \\ 1 \\ -1
  \end{pmatrix},
  \begin{pmatrix}
    0 \\ \vdots \\ 0 \\ 0 \\ 0 \\ 1
  \end{pmatrix},
\end{align*}
they form all the simple roots of $\so(d, \mathbb{C})$. Therefore $\mathfrak{g}_d \cong \so(d, \mathbb{C})$.  
\end{proof}

\section{Some integrals over \texorpdfstring{$\SU(N)$}{SU(N)}}\label{app:U-matrix-elements}
In this appendix, we evaluate some integrals over $\SU(N)$ using the invariance of the Haar measure rather than actually doing any integration explicitly, which will be useful to determine the matrix elements of $U_{ij}$ in the main text. Let's again formulate the results in the following theorem,
\begin{Theorem}\label{thm:U}
  Let $D^f (g)$ be an $N \times N$ matrix corresponding the fundamental representation of $g \in \SU(N)$. Let $\d g$ be the Haar measure over the group $\SU(N)$. Then we have the following identities
  \begin{align}
    \int \d g D^f_{ij}(g)^* D^f_{kl}(g) &= \frac{1}{N}\delta_{ik}\delta_{jl}, \label{eq:int-D*D} \\
    \int \d g D^f_{i_i j_1}(g) D^f_{i_2 j_2}(g) \cdots D^f_{i_N j_N}(g)  &= \frac{1}{N!}\varepsilon_{i_1 i_2\cdots i_N} \varepsilon_{j_1 j_2\cdots j_N}. \label{eq:int-DDD}
  \end{align}
\end{Theorem}

\begin{proof}
The key observation in the proof is that the permutation matrices are in the group $\O(N)$. In particular, all even permutations are in the group $\SO(N)$ and hence $\SU(N)$, while for odd permutations, upon changing one element from $1$ to $-1$, they are also in the group $\SU(N)$. Therefore for $\sigma$ and $\tau$ being even permutations, we have $D^f(\sigma g \tau^{-1}) = D^f(\sigma) D^f(g) D^f(\tau^{-1})$, and thus $D^f_{ij}(\sigma g \tau^{-1}) = D^f_{\sigma(i)\tau(j)}(g)$.
  
First, let's focus on the integral \cref{eq:int-D*D}. We know that $\tr D^f(g)^\dagger D^f(g) = N$ is a constant function on $\SU(N)$. Therefore
\begin{align}
  N = \int \d g \tr D^f(g)^\dagger D^f(g) = \sum_{ij}\int \d g D^f_{ji}(g)^\dagger D^f_{ij}(g) = \sum_{ij} \int \d g D^f_{ij}(g)^* D^f_{ij}(g).
\end{align}
Now for some fixed $i,j$, let's choose two even permutations $\sigma$ and $\tau$ that satisfy $\sigma(i) = 1$ and $\tau(j) = 1$. Then by changing the integration variable $g\mapsto \sigma g\tau^{-1}$, we have
\begin{align}
  \int \d g D^f_{ij}(g)^* D^f_{ij}(g) &= \int \d (\sigma g\tau^{-1}) D^f_{ij}(\sigma g\tau^{-1})^* D^f_{ij}(\sigma g\tau^{-1}) \nonumber\\
  &= \int \d (\sigma g\tau^{-1}) D^f_{\sigma(i)\tau(j)}(g)^* D^f_{\sigma(i)\tau(j)}(g) \nonumber\\
  &= \int \d g D^f_{11}(g)^* D^f_{11}(g), 
\end{align}
where in the last line we replaced $\d \sigma g\tau^{-1}$ by $\d g$ using the invariance of the Haar measure. Therefore we have
\begin{align}
  \int \d g D^f_{ij}(g)^* D^f_{ij}(g) = \frac{1}{N^2}\sum_{ij}\int \d g D^f_{ij}(g)^* D^f_{ij}(g) = \frac{1}{N}.
\end{align}
Since different matrix coefficients are orthogonal with respect to the integration over $\SU(N)$, we arrive at the result \cref{eq:int-D*D}.

Now let's prove \cref{eq:int-DDD}. Similarly we have $\det D^f(g) = 1$ is a constant function on $\SU(N)$. Therefore
\begin{align}
  1 = \int \d g \det D^f(g) = \sum_{\sigma \in S_N}\int \d g \sgn(\sigma) D^f_{1\sigma(1)}(g) \cdots D^f_{N\sigma(N)}(g) .
\end{align}
Again, we use the fact that the group $\SU(N)$ includes all even permutations. Let $\tau$ be an even permutation, and change the integration variable $g \mapsto g\tau^{-1}$, we have
\begin{align}
  \int \d g \sgn(\sigma) D^f_{1\sigma(1)}(g) \cdots D^f_{N\sigma(N)}(g) &= \int \d (g\tau^{-1}) \sgn(\sigma) D^f_{1\sigma(1)}(g\tau^{-1}) \cdots D^f_{N\sigma(N)}(g\tau^{-1}) \nonumber\\
  &= \int \d g \sgn(\sigma) D^f_{1\tau\circ\sigma(1)}(g) \cdots D^f_{N\tau\circ\sigma(N)}(g) 
\end{align}
where again we have used the invariance of the Haar measure. Then if $\sigma$ is an even permutation, we can choose $\tau\circ\sigma = (1)$, while if $\sigma$ is an odd permutation, we can choose $\tau\circ\sigma = (12)$, where we have used the cycle notation for permutations. Therefore
\begin{align}
  \int \d g (D^f_{11}(g) D^f_{22}(g) - D^f_{12}(g) D^f_{21}(g)) D^f_{33}(g) \cdots D^f_{NN}(g) = \frac{2}{N!}.
\end{align}
Now let's consider an element $h\in \SU(N)$ of the form $h = \i\sigma^2 \oplus \mathbbm{1}_{N-2}$. This element has the property that $D^f_{i2}(gh^{-1}) = D^f_{i1}(g)$ and $D^f_{i1}(gh^{-1}) = -D^f_{i2}(g)$, while all the other matrix elements $D^f_{ij}(g)$ are unchanged. Using this relation we can show that
\begin{align}
  \int \d g D^f_{11}(g) D^f_{22}(g) \cdots D^f_{NN}(g) = -\int \d g D^f_{12}(g) D^f_{21}(g) \cdots D^f_{NN}(g) = \frac{1}{N!},
\end{align}
which implies
\begin{align}
  \int \d g D^f_{1\sigma(1)}(g) \cdots D^f_{N\sigma(N)}(g) = \frac{1}{N!}\sgn(\sigma).
\end{align}
Finally, from the weight vectors of the fundamental representation and its dual representation of $\SU(N)$, we see that the decomposition of $D^f_{i_i j_1}(g) D^f_{i_2 j_2}(g) \cdots D^f_{i_N j_N}(g)$ into irreducible components contains a trivial representation only if all the $i$'s are distinct and all the $j$'s are distinct. Therefore we can write the result compactly as \cref{eq:int-DDD}.
\end{proof}
These formulas can be checked explicitly in the case of $\SU(2)$ and $\SU(3)$. A parameterization of $\SU(2)$ is given 
\begin{align}\label{eq:Df-matrix}
  D^f =
  \begin{pmatrix}
    \cos\theta \e^{\i \phi} & \sin\theta \e^{\i \psi} \\
    -\sin\theta \e^{-\i \psi} & \cos\theta \e^{-\i \phi} 
  \end{pmatrix},
\end{align}
where $\theta \in [0, \frac{\pi}{2}]$, $\phi \in [0, 2\pi)$ and $\psi \in [0, 2\pi)$, and the Haar measure on it is $\frac{1}{2\pi^2}\sin2\theta\d\theta \d\phi \d\psi$. The parameterization and Haar measure of $\SU(3)$ can be found in \cite{Bronzan:1988wa}.

Now let $|D^f\>$ be the normalized vector corresponding to $D^f$, then schematically we have the following matrix elements of $U$ when they are non-zero,
\begin{align}
  \<D^f | U |0\> &= \frac{1}{\sqrt{N}}, \\
  \<D^{f*} | U^{N-2} |D^f\> &= \pm\frac{1}{(N-1)!},
\end{align}
which are what we used in the main text.

\bibliographystyle{JHEP}
\addcontentsline{toc}{section}{References}
\bibliography{Refs,Refs-QS}

\end{document}